\documentclass{svproc}
\usepackage{url}

\usepackage[colorlinks=true, allcolors=blue]{hyperref}

\usepackage[ruled, norelsize, vlined]{algorithm2e}

\SetKwInput{kwInit}{Initialization}

\usepackage{amsmath,amsfonts,amssymb}
\usepackage{bbm,bm}
\usepackage{graphicx} 
\usepackage{tikz}
\DeclareRobustCommand\full  {\tikz[baseline=-0.6ex]\draw[thick] (0,0)--(0.5,0);}
\DeclareRobustCommand\dashed{\tikz[baseline=-0.6ex]\draw[thick,dashed] (0,0)--(0.54,0);}
\usepackage{placeins}

\newcommand{\R}{ {\mathbb{R}} }
\newcommand{\E}{ {\mathbb{E}} }
\newcommand{\Var}{ {\text{var}} }
\newcommand{\new}{ {\textsc{new}} }

\newcommand{\diag}{ {\text{diag}} }
\newcommand{\SUN}{{\textsc{sun}}}
\newcommand{\SN}{{\textsc{sn}}}
\newcommand{\EP}{{\textsc{ep}}}
\newcommand{\Bern}{{\textsc{Bern}}}

\newcommand{\br}{ {\bf r} }

\newcommand{\bx}{ {\bf x} }
\newcommand{\btilx}{ {\tilde{{\bf x}}} }
\newcommand{\bP}{ {\bf P} }

\newcommand{\bs}{ {\bf s} }
\newcommand{\bU}{ {\bf U} }
\newcommand{\by}{ {\bf y} }

\newcommand{\btilX}{ {\tilde{{\bf X}}} }

\newcommand{\bv}{ {\bf v} }

\newcommand{\bD}{ {\bf D} }
\newcommand{\bI}{ {\bf I} }
\newcommand{\bQ}{ {\bf Q} }
\newcommand{\bV}{ {\bf V} }

\newcommand{\bG}{ {\bf G} }
\newcommand{\bW}{ {\bf W} }
\newcommand{\bw}{ {\bf w} }

\newcommand{\ba}{ {\bf a} }

\newcommand{\bK}{ {\bf K} }
\newcommand{\bgamma}{ {\boldsymbol \gamma} }

\newcommand{\bvarepsilon}{ {\boldsymbol \varepsilon} }
\newcommand{\bLambda}{ {\boldsymbol \Lambda} }

\newcommand{\bGamma}{ {\boldsymbol \Gamma} }
\newcommand{\bDelta}{ {\boldsymbol \Delta} }

\newcommand{\bmu}{ {\boldsymbol \mu} }

\newcommand{\bSigma}{ {\boldsymbol \Sigma} }

\newcommand{\bOmega}{ {\boldsymbol \Omega} }
\newcommand{\bomega}{ {\boldsymbol \omega} }
\newcommand{\balpha}{ {\boldsymbol \alpha} }
\newcommand{\bxi}{ {\boldsymbol \xi} }
\newcommand{\bz}{ {\bf z} }
\newcommand{\btheta}{ {\boldsymbol \theta} }

\begin{document}
\mainmatter              
\title{Expectation propagation for the smoothing distribution in dynamic probit}
\titlerunning{EP for dynamic probit}  
%
\author{Niccolò Anceschi\inst{1}
\and
Augusto Fasano\inst{2}
\and
Giovanni Rebaudo\inst{2,3}
}

\authorrunning{N. Anceschi et al.} 
%
\tocauthor{Niccolò Anceschi, Augusto Fasano, and Giovanni Rebaudo}
\institute{Duke University, Durham, USA. \email{niccolo.anceschi@duke.edu}
\and
Collegio Carlo Alberto, Turin, IT. \email{augusto.fasano@carloalberto.org}
\and
University of Turin, Turin, IT. \email{giovanni.rebaudo@unito.it}
}

\maketitle              

\begin{abstract}
The smoothing distribution of dynamic probit models with Gaussian state dynamics was recently proved to belong to the unified skew-normal family.
Although this is computationally tractable in small-to-moderate settings, it may become computationally impractical in higher dimensions.
In this work, adapting a recent more general class of expectation propagation (\EP) algorithms, we derive an efficient \EP\ routine to perform inference for such a distribution.
We show that the proposed approximation leads to accuracy gains over available approximate algorithms in a financial illustration.

\end{abstract}
\section{Introduction}
\label{sec:1}
Dynamic binary regression is a lively field of research, with the dynamic probit model playing a central role \cite{andrieu2002particle,fasano2021closed,fasano2021variational,he2023dynamic}.
In this work, we develop an expectation propagation (\EP) approximation of the joint smoothing distribution in a dynamic probit model. 
In such a state-space model, for each time $t$, a known $p$-dimensional covariate vector $\bx_t$ is available, while the  binary observations $y_t\in\{0;1\}$, $t=1,\ldots,n$, compose a time series whose dependence across time is driven by a dynamic latent state $\btheta_t=(\theta_{1t},\ldots,\theta_{pt})^\intercal\in\R^p$. The latter has Markovian dynamics and affects the probability of success of the observation at the corresponding time according to the following specifications:
\begin{eqnarray}
y_{t} \mid \btheta_{t} &&\sim \Bern (\Phi(\bx_t^\intercal\btheta_{t})), \label{eq1}\\ 
\btheta_t &&=\bG_{t}\btheta_{t-1}+\bvarepsilon_t, \quad \bvarepsilon_t \overset{\text{ind}}{\sim}  \mbox{N}_p({\bf 0}, \bW_t), \quad t=1 \ldots, n, \label{eq2}
\end{eqnarray}
with $\btheta_0 \sim \mbox{N}_p(\ba_0, \bP_0) \perp \{\bvarepsilon_t \}_{t\ge 1}$.
We denote with $\Phi(\cdot)$ the cumulative distribution function of the standard normal distribution, while $\bG_t$, $\bW_t$ and $\bP_0$ are known matrices. 
In the following, according to common practices, we set $\ba_0 = \boldsymbol{0}$.
Moreover, we will denote with $\Phi_m(\bz;\bSigma)$ the cumulative distribution of an $m$-variate Gaussian random variable having zero mean and covariance matrix $\bSigma$, evaluated at $\bz$.
Calling $\by_{1:n}=(y_1,\ldots,y_n)^\intercal$ and $\btheta_{1:n}=(\btheta_1^\intercal,\ldots,\btheta_n^\intercal)^\intercal$, the smoothing distribution is given by $p(\btheta_{1:n}\mid \by_{1:n})$, which has been showed (see \cite{fasano2021closed}) to belong to the class of unified skew-normal (\SUN) distributions \cite{arellano2006unification}.
This result comes however with some computational difficulties when the length of the time series is moderately large, for the reasons clarified in Section~\ref{sec:2}, motivating the development of approximate methods \cite{fasano2021variational}.
This can be done by resorting to expectation propagation, which constructs a Gaussian approximation of the true posterior mirroring its factorization structure \cite{Vehtari2020ep}.
To this end, it is worth noting that the model \eqref{eq1}-\eqref{eq2} can be rewritten as a probit model with parameter $\btheta_{1:n}\in \R^{pn}$ and fictional $pn$-dimensional covariates $\btilx_t=(\boldsymbol{0}_{p(t-1)}^\intercal,\bx_t^\intercal,\boldsymbol{0}_{p(n-t)}^\intercal)^\intercal$, with $\boldsymbol{0}_m$ denoting the $m$-dimensional column vector of zeros.
Indeed, model \eqref{eq1}-\eqref{eq2} is equivalent to \vspace{-7pt}
\begin{align}
    p(y_{t} \mid \btheta_{1:n}) &= \Phi((2y_t-1)\btilx_t^\intercal\btheta_{1:n}), \label{eq3}\\ 
\btheta_{1:n} &\sim N_{pn}(\boldsymbol{0},\bOmega) \label{eq4}, \vspace{-10pt}
\end{align}
where $\bOmega$ is composed by $(p \times p)$-dimensional blocks whose expression follows directly from the dynamics in \eqref{eq2}: defining $\bG_{l}^{t}=\bG_t \cdots \bG_l$, $l\le t-1$, it holds $\bOmega_{[tt]}=\mbox{var}(\btheta_t)=\bG^{t}_1 \bP_0 \bG^{t\intercal}_1+\sum_{l=2}^t\bG^{t}_l \bW_{l-1}\bG^{t\intercal}_l+\bW_t$, for $t=1, \ldots,n$, and  $\bOmega_{[tl]}=\bOmega^{\intercal}_{[lt]}=\mbox{cov}(\btheta_t, \btheta_l)=\bG_{l+1}^{t}\bOmega_{[ll]}$, for $t>l$ (see also \cite{fasano2021variational}).
From this,\vspace{-5pt}
\begin{equation}
\label{eq5}
    p(\btheta_{1:n}\mid \by_{1:n})\propto p(\btheta_{1:n}) \prod_{t=1}^n \Phi((2y_t-1)\btilx_t^\intercal\btheta_{1:n}).\vspace{-5pt}
\end{equation}
Leveraging on this representation, the \EP\ scheme constructs an  approximating density of \eqref{eq5} by replacing each factor therein with a Gaussian density, whose parameters are optimized iteratively via moments matching between the global approximant and a so-called hybrid distribution, as detailed in Section~\ref{sec:3}.
The resulting global \EP\ approximation, which is Gaussian by construction, is shown to be accurate and tractable in Section~\ref{sec:4}.
More details on the \EP\ algorithm can be found, for instance, in Chapter 10 of \cite{bishop2006pattern}.

\section{Literature review}
\label{sec:2}

Exploiting \eqref{eq5}, \cite{fasano2021closed} extended the conjugacy results in the static probit regression \cite{durante2019conjugate} to the more challenging multivariate dynamic probit setting.
In particular, \cite{fasano2021closed} showed that the filtering, predictive, and smoothing densities of the state variables in this setting have unified skew-normal (\SUN) \cite{arellano2006unification} kernels and compute the closed form of the parameters of such distributions.
To provide a brief overview, a random vector $\btheta \in \R^q$ is said to follow a \SUN\ distribution, denoted as $\btheta \sim \mbox{SUN}_{q,h}(\bxi,\bOmega,\bDelta,\bgamma,\bGamma)$, if its density function can be expressed as:
\begin{equation*}
\phi_q(\btheta -\bxi; \bOmega) \frac{\Phi_h\left(\bgamma+\bDelta^\intercal \bar{\bOmega}^{-1} \bomega^{-1}(\btheta-\bxi); \bGamma{-}\bDelta^{\intercal}\bar{\bOmega}^{-1}\bDelta \right)}{\Phi_h(\bgamma;\bGamma)},
\end{equation*}
where $\phi_q(\btheta -\bxi; \bOmega)$ is the density function of a Gaussian distribution with $\boldsymbol{0}$ mean and covariance matrix $\bOmega$ evaluated at $\btheta -\bxi$, $\bomega=(\bOmega \odot \mathbf{I}_q)^{1/2}$ is the diagonal scale matrix, with $\odot$ denoting the element-wise Hadamard product, and $\bar{\bOmega}=\bomega^{-1}\bOmega\bomega^{-1}$ is the corresponding correlation matrix.
Additional details on \SUN\ distributions can be found in \cite{arellano2006unification}.
More specifically, taking $\bOmega$ as in Section~\ref{sec:1}, under model \eqref{eq1}--\eqref{eq2} the joint smoothing distribution has the form (see \cite{fasano2021closed,fasano2021variational})
\begin{equation}
	(\btheta_{1:n} \mid \by_{1:n}) \sim \mbox{\textsc{sun}}_{pn, n}(\boldsymbol{0},\bOmega_{1:n\mid n}, \bDelta_{1:n\mid n},  \boldsymbol{0}, \bGamma_{1:n\mid n}), 
 \label{eq:JointSmoothing}
\end{equation}
with $\bOmega_{1:n\mid n}=\bOmega$, $\bDelta_{1:n\mid n}=\bar{\bOmega}\bomega\bD^{\intercal}\bs^{-1}$, $\bGamma_{1:n\mid n}=\bs^{-1}(\bD\bOmega\bD^{\intercal}+\bI_n)\bs^{-1}$, where $\bD$ is an $n\times pn$ block-diagonal matrix having block entries $\bD_{[tt]} = (2y_t-1)\bx_t^\intercal$, $t=1,\ldots,n$,  $\bs=[(\bD\bOmega\bD^{\intercal}+\bI_n) \odot \mbox{\bf I}_{n}]^{1/2}$, $\bI_n$ defines the $n$-dimensional identity matrix.
Leveraging on \eqref{eq:JointSmoothing} and the additive representation of the \textsc{sun} \cite{arellano2006unification}, one can establish a probabilistic characterization that facilitates the generation of independent and identically distributed (i.i.d.) samples from the smoothing distribution. 
The characterization is as follows:
\begin{equation*}
(\btheta_{1:n} \mid \by_{1:n}) \stackrel{\text{d}}{=} \bomega_{1:n\mid n}(\bU_{0 \ 1:n\mid n}+\bDelta_{1:n\mid n} \bGamma_{1:n\mid n}^{-1} \bU_{1 \ 1:n\mid n}),
\end{equation*}
where $\bU_{0 \ 1:n\mid n} \sim \mbox{N}_{pn}({\bf 0},\bar{\bOmega}_{1:n\mid n}- \bDelta_{1:n\mid n}\bGamma_{1:n\mid n}^{-1}\bDelta_{1:n\mid n}^{\intercal})$, while $\bU_{1 \ 1:n\mid n}$ is distributed according to a truncated $\mbox{N}_{n}({\bf 0},\bGamma_{1:n\mid n})$ with lower truncation at $\boldsymbol{0}$.
Based on this representation, one can develop an i.i.d.\ sampler where the most computationally intensive task is sampling from an $n$-variate truncated Gaussian distribution. 
However, despite recent advances \cite{botev2017normal} enabling efficient simulation for small-to-moderate time series (i.e., with $n$ in the order of a few hundred), this i.i.d.\ sampler may become impractical for longer time series due to such computational constraints arising from the multivariate truncated Gaussian component.

To overcome this issue, \cite{fasano2021variational} developed a partially-factorized variational Bayes (\textsc{pfm-vb}) approximation of the smoothing distribution where, adapting \cite{fasano2022scalable} to the dynamic setting, the multivariate truncated Gaussian component is replaced by $n$ independent univariate truncated Gaussian terms.
This leads to remarkable computational gains and great accuracy.
Here, motivated by the overall improved accuracy of \EP\ \cite{anceschi2023bayesian,chopin2017leave},
we propose an efficient and accurate \EP\ algorithm for the smoothing distribution of model \eqref{eq1}-\eqref{eq2}. 
Our approach adapts to the dynamic probit setting techniques developed by \cite{anceschi2023bayesian} and further specified to the static probit in \cite{fasano2023efficient}.

\section{Expectation propagation (EP) for the dynamic probit}
\label{sec:3}
As clarified in Section \ref{sec:2}, dealing with the $\SUN_{pn,n}$ smoothing distribution for model \eqref{eq1}-\eqref{eq2} may become computationally intractable in scenarios where $n$ is not small-to-moderate.
As anticipated, the computational issues arise from the fact that sampling from such distribution requires sampling from an $n$-variate truncated normal and similar problems are faced when computing moments.
Consistently, computational methods which are able to provide a good approximation of the smoothing distribution at a much lower computational time may provide a preferable alternative.
In line with this goal, we derive the expectation propagation (\EP) routine \cite{bishop2006pattern,minka2001expectation} for the dynamic probit model \eqref{eq1}-\eqref{eq2}, by specifying more general results obtained for a wide class of models in \cite{anceschi2023bayesian} and adapting computations done for the static probit model in \cite{fasano2023efficient}.
Section \ref{subsec:3.1} presents how one can leverage results on multivariate extended skew-normal distributions (\SN) to obtain the equations for the \EP\ updates.
The resulting approximation is computable at a much lower cost than sampling from the exact smoothing distribution.
Nevertheless, it requires matrices manipulation of dimension $pn\times pn$ within each iteration, which may be computationally inefficient.
We thus show in Section \ref{subsec:3.2} how the \EP\ routine can be implemented avoiding such direct matrix computations, by storing and updating only lower-dimensional parameters.
This results in sensible computational gains.

\subsection{Implementation without $pn\times pn$ matrix inversions}
\label{subsec:3.1}
The goal of expectation propagation (\EP) is to approximate the smoothing distribution $p(\btheta_{1:n}\mid \by_{1:n})$ with an approximating distribution having density $q(\btheta_{1:n})\propto\prod_{t=0}^n q_t(\btheta_{1:n})$, whose factorization reflects the form of the smoothing distribution \eqref{eq5}.
Differently from \eqref{eq5}, in order to have a tractable approximating distribution, all the $q_t$'s, $t=0,\ldots,n$, are Gaussian densities having form $q_t(\btheta_{1:n})\propto\exp\left(-\frac{1}{2}\btheta_{1:n}^\intercal\bQ_t\btheta_{1:n} +\btheta_{1:n}^\intercal\br_t\right)$ for $t=1,\ldots,n$.
As a consequence, the resulting \EP\ approximating density will be Gaussian too, with $q(\btheta_{1:n}) = \phi_p(\btheta_{1:n}-\bQ^{-1}\br,\bQ^{-1})$, having set $\br=\sum_{t=0}^n \br_t$, $\bQ = \sum_{t=0}^n\bQ_t$.
Recalling the analogy between the factorization of $q(\btheta_{1:n})$ and \eqref{eq5}, the parameters for the factor, or site, $0$ are set to $\br_0=\boldsymbol{0}$ and $\bQ_0=\bOmega^{-1}$ and kept constant throughout the algorithm, so that $q_0(\btheta_{1:n})$ matches the prior distribution.
On the other hand, the parameters $\br_t$ and $\bQ_t$ for each site $t=1,\ldots,n$ are iteratively updated so that, keeping the parameters for the other sites fixed, the first two moments of the global approximation $q(\btheta_{1:n})$ match the ones of the hybrid distribution
\begin{equation}
	h_t(\btheta_{1:n}) \propto p(y_t\mid \btheta_{1:n}) \prod_{j\ne t} q_j(\btheta_{1:n}) =
	\Phi((2y_t-1)\btilx_t^\intercal\btheta_{1:n})\prod_{j\ne t} q_j(\btheta_{1:n}).
\end{equation}
The moments of $q(\btheta_{1:n})$ are immediate, being $q(\btheta_{1:n})$ Gaussian, while the ones of $h_t(\btheta_{1:n})$ are also straightforward to compute, after noticing that it coincides with the kernel of an $\SN_p(\bxi_t,\bOmega_t,\balpha_t,\tau_t)$ \cite{azzalini2014skew} with
\begin{equation*}
\begin{split}
    \bxi_t &= \bQ_{-t}^{-1}\br_{-t},\quad 
	\bOmega_t = \bQ_{-t}^{-1},\quad
    \balpha_t = (2y_t-1)\bomega_t\btilx_t,\\
	\tau_t &= (2y_t-1)(1+\btilx_t^\intercal\bOmega_t\btilx_t)^{-1/2}\btilx_t^\intercal\bxi_t,
\end{split}
\end{equation*}
with $\bQ_{-t}=\sum_{j\ne t} \bQ_j=\bQ-\bQ_t$, $\br_{-t}=\sum_{j\ne t}\br_j=\br-\br_t$ and $\bomega_t=\left(\bOmega_t\odot \bI_{pn}\right)^{1/2}$.
The moments of such distribution are then given by Equations (5.71) and (5.72) in \cite{azzalini2014skew}. Calling $\zeta_1(x) = \phi(x)/\Phi(x)$ and $\zeta_2(x)=-\zeta_1(x)^2-x\zeta_1(x)$, they equal
\begin{equation*}
\begin{split}
	\bmu_{h_t} &= \E_{h_t(\btheta_{1:n})} [\btheta_{1:n}] = \bxi_t+\zeta_1(\tau_t) s_t\bOmega_t\btilx_t\\
	\bSigma_{h_t} &= \Var_{h_t(\btheta_{1:n})} [\btheta_{1:n}] = \bOmega_t + \zeta_2(\tau_t) s_t^2\bOmega_t\btilx_t\btilx_t^\intercal\bOmega_t,
\end{split}
\end{equation*}
with $s_t=(2y_t-1)(1+\btilx_t^\intercal\bOmega_t\btilx_t)^{-1/2}$.
In \EP, at each iteration the parameters for $\br_t$ and $\bQ_t$ of each site $t=1,\ldots,n$ are updated so that the \EP\ moment matching conditions
\begin{equation*}
	\begin{cases}
	\left(\bQ_{-t}+\bQ_t^\new\right)^{-1}(\br_{-t} +\br_t^\new) = \bmu_{h_t}\\
	\left(\bQ_{-t}+\bQ_t^\new\right)^{-1} = \bSigma_{h_t},
	\end{cases}
\end{equation*}
are satisfied by the updated parameters $\br_t^\new$ and $\bQ_t^\new$.
This translates into
\begin{equation*}
	\begin{cases}
	\br_t^\new = \left(\bQ_{-t}+\bQ_t^\new\right) \bmu_{h_t} - \br_{-t}\\
	\bQ_t^\new = \bSigma_{h_t}^{-1} - \bQ_{-t}.
	\end{cases}
\end{equation*}
\begin{algorithm}[t]
 \caption{\EP\ for dynamic probit - no $pn\times pn$ matrix inversions}
 \label{algo1}
\kwInit{\mbox{$\bQ^{-1} = \bOmega;\,\ \br=\boldsymbol{0}$; $\,\ k_t = 0$ and $m_t = 0$ for $t=1,\ldots,n$.}}
 \For{$\, s \,$ from $\, 1 \,$ until convergence $ $}{
 \For{$\, t \,$ from $\, 1 \,$ to $\, n \,$}{
 $\br_{-t} = \br - m_t \btilx_t $\\ 
 $\bOmega_t = \bQ^{-1} + k_t/\left(1- k_t \btilx_t^\intercal \bQ^{-1} \btilx_t\right) \bQ^{-1} \btilx_t \btilx_t^\intercal
 \bQ^{-1}$ \\
 $s_t = (2y_t-1)(1+\btilx_t^\intercal \bOmega_t \btilx_t)^{-1/2}$ \\
 $\tau_t = s_t \btilx_t^\intercal \bOmega_t \; \br_{-t} $\\[2pt]
 $k_t = -\zeta_2(\tau_t)/\left(1 + \btilx_t^\intercal\bOmega_t\btilx_t + \zeta_2(\tau_t)\btilx_t^\intercal\bOmega_t\btilx_t\right)$ \\[2pt]
 $m_t = \zeta_1(\tau_t) s_t + k_t \btilx_t^\intercal \bOmega_t\br_{-t} + k_t \zeta_1(\tau_t) s_t \btilx_t^\intercal \bOmega_t\btilx_t $\\[2pt]
 $\br = \br_{-t} + m_t \btilx_t $\\
 $\bQ^{-1}=\bOmega_t + \zeta_2(\tau_t) s_t^2\bOmega_t\btilx_t\btilx_t^\intercal\bOmega_t$
 }
 }
\KwOut{$ q(\btheta_{1:n})=\phi_p(\btheta_{1:n} - \bQ^{-1}\br; \bQ^{-1}) $}
\end{algorithm}
Adapting the derivations in \cite{fasano2023efficient}, where results for a broader class of models reported in \cite{anceschi2023bayesian} are specified for the classical probit model, the application of Woodbury's identity to  $\bSigma_{h_t}^{-1}$ and some algebra lead to the equalities
\begin{equation*}
    \bQ_t^\new = k_t^\new \btilx_t \btilx_t^\intercal,\quad \br_t^\new = m_t^\new \btilx_t,
\end{equation*}
with $k_t^\new = -\zeta_2(\tau_t)/\left(1 + \btilx_t^\intercal\bOmega_t\btilx_t + \zeta_2(\tau_t)\btilx_t^\intercal\bOmega_t\btilx_t\right)\ $
and $m_t^\new = \zeta_1(\tau_t) s_t + k_t^\new   \btilx_t^\intercal \bOmega_t \br_{-t} + k_t^\new \zeta_1(\tau_t) s_t \btilx_t^\intercal \bOmega_t\btilx_t$.
Thus, the updates of the multidimensional parameters $\br_t$ and $\bQ_t$ are fully determined by the updates of the scalar quantities $k_t$ and $m_t$, $t=1,\ldots,n$.
These are usually initialized to zero so that the initial \EP\ approximation $q(\btheta_{1:n})$ coincides with the prior distribution for $\btheta_{1:n}$.
In order to be able to implement the updates of $k_t$ and $m_t$, $t=1,\ldots,n$, one needs to compute $\bOmega_t=\bQ_{-t}^{-1}$.
This can be done by avoiding direct matrix inversions during the iterations, exploiting Woodbury's identity. Indeed, it holds
\begin{equation*}
 \bOmega_t  = \bQ_{-t}^{-1} = \left(\bQ - k_t \btilx_t \btilx_t^\intercal \right)^{-1}
    =  \bQ^{-1} + \dfrac{k_t}{1- k_t \bx_t^\intercal \bQ^{-1} \btilx_t} \bQ^{-1} \btilx_t\btilx_t^\intercal \bQ^{-1} ,
\end{equation*}
with $\bQ^{-1}$ computed explicitly in the precomputation step and then updated at each step after the site-specific parameters are computed, exploiting the equality $(\bQ^\new)^{-1}=(\bQ_{-t}+\bQ_t^{\new})^{-1}=\bSigma_{h_t}=\bOmega_t + \zeta_2(\tau_t) s_t^2\bOmega_t\btilx_t\btilx_t^\intercal\bOmega_t$.
The results presented above form the \EP\ implementation presented in Algorithm \ref{algo1}.
Such implementation does not involve direct matrix inversions across iterations.
Nevertheless, it requires storing, updating and multiplying the $pn\times pn$-dimensional matrices $\bOmega_t$, $t=1,\ldots,n$, and $\bQ^{-1}$.
This may result inefficient when either $p$ or $n$ (or both) are large.
We then show in Section \ref{subsec:3.2} how the \EP\ routine can be implemented without storing and manipulating such $pn\times pn$-dimensional matrices, but only lower dimensional ones.

\subsection{Implementation without $pn\times pn$ matrix updates}
\label{subsec:3.2}
\begin{algorithm}[b!]
    \label{algo2}
    \caption{\mbox{\EP\ for dynamic probit - no $pn\times pn$ 
 matrix inversions or updates}}
 \kwInit{\mbox{$\br=\boldsymbol{0}$; $\,\ k_t = 0$ and $m_t = 0$ for $t=1,\ldots,n$; $\,\ \bV=\left[\bv_1,\ldots,\bv_n \right]=\bOmega\btilX^\intercal$.}} 
 \For{$\, s \,$ from $\, 1 \,$ until convergence $ $}{
 \For{$\, t \,$ from $\, 1 \,$ to $\, n\,$}{
 $\bw_t = (1-k_t \btilx_t^\intercal \bv_t)^{-1} \bv_t $ \\
 $\br_{-t} = \br - m_t \btilx_t $\\[2pt]
 $s_t = (2y_t-1)(1+\btilx_t^\intercal \bw_t)^{-1/2}$ \\
 $\tau_t = s_t \bw_t^\intercal \; \br_{-t} $\\[2pt]
 $k_t^\new = -\zeta_2(\tau_t)/\left(1 + \btilx_t^\intercal\bw_t + \zeta_2(\tau_t)\btilx_t^\intercal\bw_t\right)$\\[2pt]
 $m_t = \zeta_1(\tau_t) s_t + k_t^\new \bw_t^\intercal \br_{-t} + k_t^\new \zeta_1(\tau_t) s_t \btilx_t^\intercal \bw_t $\\[2pt]
 $k_t = k_t^{\text{new}}$\\
 $\br = \br_{-t} + m_t \btilx_t $ \\
 $\bV = \bV - \bv_t \left[ (k_t^\new - k_t)/ \left(1 + (k_t^\new - k_t) \btilx_t^\intercal \bv_t\right) \right] \btilx_t^\intercal \bV$
 }
 }
$\bQ^{-1}=\bOmega-\bV\bK\btilX\bOmega $\\[2pt]
\KwOut{$ q(\btheta_{1:n})=\phi_p(\btheta_{1:n} - \bQ^{-1}\br; \bQ^{-1}) $}
\end{algorithm}

As shown in Section \ref{subsec:3.1}, the parameters $\br_t$ and $\bQ_t$, $t=1,\ldots,n$, are fully characterized by the known vector $\btilx_t$ and the scalar quantities $k_t$ and $m_t$, which are iteratively updated in the \EP\ algorithm.
A close inspection of Algorithm \ref{algo1} shows that the quantities $\bQ^{-1}$ and $\bOmega_t$, $t=1,\ldots,n$, are not explicitly needed to perform such updates, but it is enough to store and update the $pn$-dimensional vectors $\bw_t = \bOmega_t \btilx_t = \bQ_{-t}^{-1}\btilx_t$ and $\bv_t = \bQ^{-1}\btilx_t$, $t=1,\ldots,n$.
Thus, in order to be able to implement the \EP\ routine in terms of these quantities, their update rules need to be derived and, most importantly, they must be computable in an efficient way.
We show here that they allow an efficient formulation, which may make the new implementation preferable, from a computational point of view, to Algorithm \ref{algo1}.
Adapting the derivations in \cite{fasano2023efficient}, applying Woodbury's identity to $\bQ_{-t}^{-1}=(\bQ-\bQ_t)^{-1}$, after some algebra one gets $\bw_t=d_t \bv_t$, with $d_t=(1-k_t\btilx_t^\intercal\bv_t)^{-1}$.
Thus, when updating the parameters for site $t$, $t=1,\ldots,n$, $\bw_t$ can be computed immediately from $\bv_t$.
On the other hand, each time a site $t$ is updated also the \EP\ covariance matrix $\bQ^{-1}$ changes.
Thus, at the end of the update of each site $t$, \textit{all} the $\bv_j$'s (not only $\bv_t$) must be updated to reflect this change.
After some algebra, exploiting again Woodbury's identity, one gets
\begin{equation*}
\begin{split}
\bv_j^\new &= (\bQ^{\new})^{-1} \btilx_j = (\bQ - \bQ_t+\bQ_t^\new)^{-1}\btilx_j
= [\bQ + (k_t^\new - k_t) \btilx_t \btilx_t^\intercal]^{-1}\btilx_j\\
&= [\bQ^{-1} - (k_t^\new - k_t) [1+(k_t^\new - k_t)\btilx_t^\intercal\bQ^{-1}\btilx_t ]^{-1} \bQ^{-1}\btilx_t\btilx_t^{\intercal}\bQ^{-1} ]\btilx_j\\
&= \bQ^{-1}\btilx_j-[(k_t^\new - k_t)^{-1} +\btilx_t^\intercal\bv_t ]^{-1} \bv_t\btilx_t^\intercal\bv_j
= \bv_j - c_t (\btilx_t^\intercal\bv_j) \bv_t,
\end{split}
\end{equation*}
where $c_t=(k_t^\new - k_t)/ (1 + (k_t^\new - k_t)\btilx_t^\intercal\bv_t)$.
Storing the $\bv_j$'s in the $pn\times n$ matrix $\bV=[\bv_1,\bv_2,\dots,\bv_n]$, these updates translate in the matrix update $\bV^\new = \bV - c_t \bv_t \btilx_t^\intercal \bV$.
Finally, after the algorithm has converged, one has to  derive the form of the \EP\ covariance matrix $\bQ^{-1}$ from the quantities used in the updates.
It holds $\bQ^{-1} =\big(\sum_{t=0}^n\bQ_t\big)^{-1} =\big(\bOmega^{-1}+\sum_{t=1}^n\bQ_t\big)^{-1}=\big(\bOmega^{-1} + \btilX^\intercal \bK \btilX\big)^{-1}$, having defined  $\btilX=(\btilx_1,\ldots,\btilx_n)^\intercal$ and $\bK = \diag(k_1,\ldots,k_n)$.
Consequently, calling $\bLambda=(\bI_n+\bK\btilX\bOmega\btilX^\intercal)^{-1}$, exploiting Woodbury's identity, it trivially follows that $\bQ^{-1}=\bOmega-\bOmega\btilX^\intercal\bLambda\bK\btilX\bOmega$.
From this, one has $\bV=\bQ^{-1}\btilX^\intercal=\bOmega\btilX^\intercal[\bI_n-\bLambda\bK\btilX\bOmega\btilX^\intercal]=\bOmega\btilX^\intercal\bLambda[\bLambda^{-1}-\bK\btilX\bOmega\btilX^\intercal]=\bOmega\btilX^\intercal\bLambda$
so that $\bQ^{-1}=\bOmega-\bV\bK\btilX\bOmega$, showing that $\bQ^{-1}$ can be computed without any direct matrix inversion.
Algorithm \ref{algo2} reports all the steps needed to perform such efficient \EP\ implementation.\\

\subsection{Computational costs}
\label{subsec:3.3}

Calling $\tilde{p}$ the total number of parameters, \cite{anceschi2023bayesian} and \cite{fasano2023efficient} noted that, in the case of static probit regression, the cost of each \EP\ update for the corresponding versions of Algorithm \ref{algo1} and \ref{algo2} scales as $O(n \tilde{p}^2)$ and $O(n^2 \tilde{p})$, respectively.
Adapting those derivations to the current dynamic setting, where $\tilde{p}=pn$, one obtains that the cost of each \EP\ update for Algorithms \ref{algo1} and \ref{algo2} scales respectively as $O(n^3 p^2)$ and $O(n^3 p)$, making the latter always more efficient, differently from the static case, where Algorithm~\ref{algo2}  becomes advantageous only in high-dimensional scenarios, with $p > n$.
The same consideration holds true even accounting for operations required to initialize and conclude the main \EP\ routines.
Calling $N_{\EP}$ the number of iterations needed to reach convergence, it is straightforward to show that Algorithms \ref{algo1} and \ref{algo2} scale respectively as $O(n^3 p^2 N_{\EP}) = O(n^3 p (p \cdot N_{\EP}))$ and $O(n^3 p (p+N_{\EP}))$.
This result makes our contribution even more significant, as the proposed \EP\ implementation leads to computational improvements regardless of being applied to high-dimensional data $\bx_t$ or not.
Finally, it is worth mentioning that the sparse nature of  $\btilx_t$ allows us to further reduce the computational costs of different steps in Algorithms \ref{algo1} and \ref{algo2}.
Albeit reducing the number of maximum-cost operations, this does not alter the overall scaling of the \EP\ routines.

\section{Financial illustration}
\label{sec:4}
We demonstrate the performance of the \EP\ approximation, derived in Section~\ref{sec:3} in a financial application that was previously used in \cite{fasano2021variational} for comparisons of some state-of-the-art algorithms to approximate the smoothing distribution \eqref{eq:JointSmoothing}. 
More precisely, we consider a dynamic probit regression for the daily opening directions of the French \textsc{cac}40 stock market index, spanning from January 4th, 2018, to December 28th, 2018, comprising a total of $n=241$ observations.
In this study, the binary response variable $y_t$ is defined as $1$ if the opening value of the \textsc{cac}40 on the day $t$ exceeds the corresponding closing value from the previous day and $0$ otherwise.
We incorporate two covariates: the intercept and the opening direction of the \textsc{nikkei}225, treated as a binary covariate and denoted as $\xi_t$. 
As the Japanese market opens before the French one, $\xi_t$ is available prior to $y_t$, making it a valid predictor for each day $t$. 
Therefore, in reference to model \eqref{eq1}-\eqref{eq2}, we have $p=2$ and $\bx_t=(1, \xi_t)^\intercal$.
Additionally, we set $\bW_t = \text{diag}(0.01,0.01)$ for all $t$ and $\bP_0=\text{diag}(3,3)$. 
Detailed information about the hyperparameters' setting can be found in \cite{fasano2021closed}.

\begin{figure}[t!]
    \centering  
    \includegraphics[width=\linewidth]{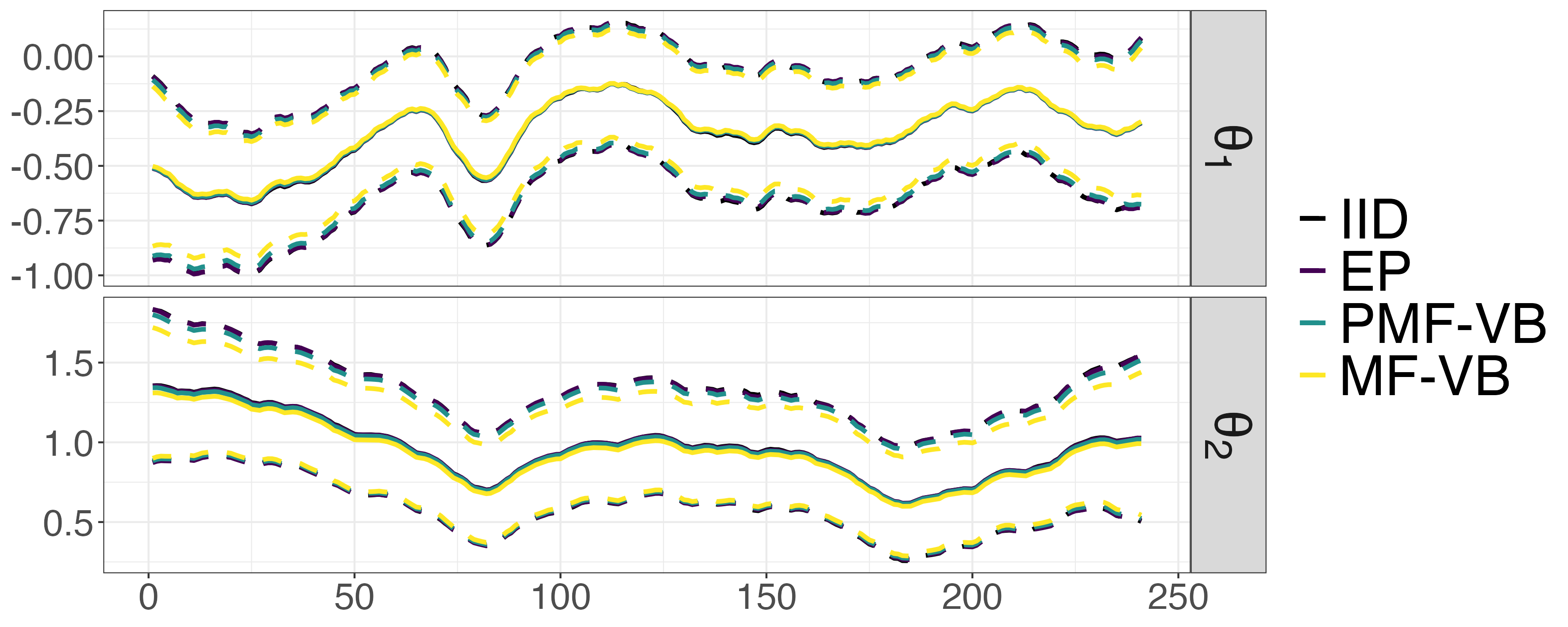}
    \caption{$\mathbb{E}[\btheta_{1:n}\mid\by_{1:n}]$ (\full) and $\mathbb{E}[\btheta_{1:n}\mid\by_{1:n}] \pm \sqrt{\text{var}[\btheta_{1:n}\mid\by_{1:n}]}$ (\dashed) for the i.i.d.\ sampler (\textsc{iid}) and the \textsc{ep}, \textsc{pfm-vb} and \textsc{mf-vb} approximations.}     
    \label{fig:Plot_smooth}
\end{figure}

\begin{figure}[t]
    \centering  \includegraphics[width=0.9\linewidth]{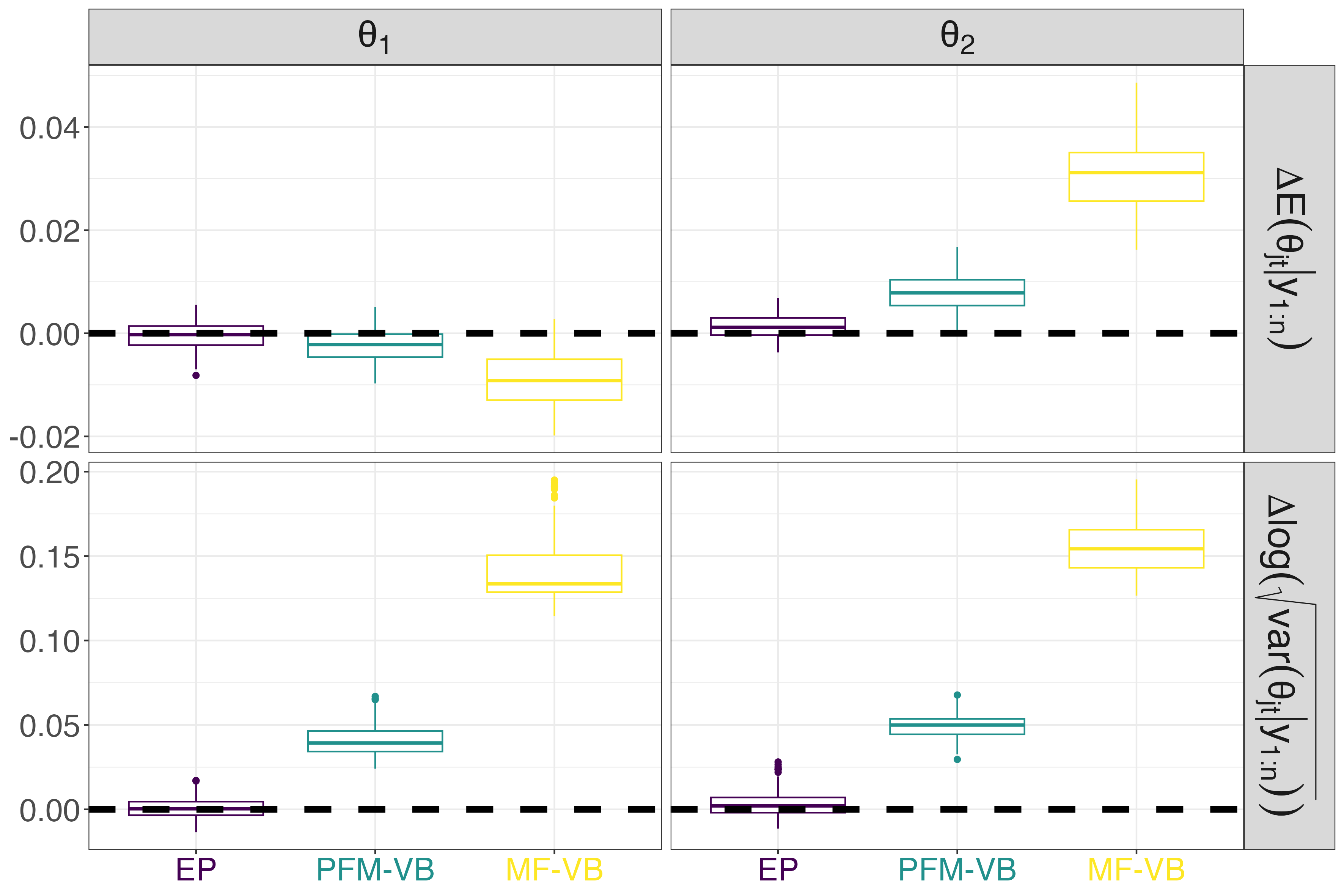}
    \caption{Boxplot of the differences of the $\E[\theta_{j1}\mid\by_{1:n}], \ldots, \E[\theta_{jn}\mid\by_{1:n}]$ and $\log\left(\sqrt{\text{var}[\theta_{j1}\mid\by_{1:n}]}\right), \ldots, \log\left(\sqrt{\text{var}[\theta_{jn}\mid\by_{1:n}]}\right)$, $j=1,2$, obtained with the \EP\, the \textsc{pfm-vb}, and the \textsc{mf-vb} solutions, using the inferences obtained via i.i.d.\ sampling from the exact \textsc{sun} as benchmark.}     
    \label{fig:Plot_Diff}
\end{figure}

Inference about the smoothing distribution \eqref{eq:JointSmoothing} is conducted using as benchmark the results arising from $10^4$ samples from the i.i.d.\ sampler in \cite{fasano2021closed}.
We then compare the accuracy of three approximate methods in recovering functionals of interest of the smoothing distribution.
More precisely, we consider the \EP\ approximation implemented as in Algorithm~\ref{algo2}, the \textsc{pfm-vb} introduced in \cite{fasano2021variational} briefly described in Section~\ref{sec:2}, and a mean-field variational Bayes (\textsc{mf-vb}) approximation which adapts \cite{consonni2007mean} to the dynamic setting.
These results are illustrated in Figure \ref{fig:Plot_smooth}, where we plot $\mathbb{E}[\btheta_{1:n}\mid\by_{1:n}]$ and the bands $\mathbb{E}[\btheta_{1:n}\mid\by_{1:n}] \pm \sqrt{\text{var}[\btheta_{1:n}\mid\by_{1:n}]}$.
It can be seen that, although all approximate methods reach a reasonable accuracy, \textsc{mf-vb} shows some over-shrinkage of posterior moments towards zero, while \EP\ is slightly more precise than \textsc{pfm-vb}.
This is made clearer in Figure~\ref{fig:Plot_Diff}, where we show the boxplots of the differences between the smoothing means $\E[\theta_{j1}\mid\by_{1:n}],\ldots,\E[\theta_{jn}\mid\by_{1:n}]$ and log standard deviations $\log\left(\sqrt{\text{var}[\theta_{j1}\mid\by_{1:n}]}\right),\ldots,\log\left(\sqrt{\text{var}[\theta_{jn}\mid\by_{1:n}]}\right)$, $j=1,2$, obtained via the i.i.d.\ sampler and the ones resulting from the approximate methods.
From there, the above-mentioned over-shrinkage of the \textsc{mf-vb} approximation is immediate to notice, while the improvements of \EP\ over \textsc{pfm-vb} are also clarified.
Finally, we note that the approximated methods enable the computation of the desired moments in less than half a second (on a MacBook Pro 14-inch, 2023 the \EP\ takes $0.43$ secs, the \textsc{pfm-vb} takes $0.27$ secs, and the \textsc{mf-vb} takes $0.20$ secs). 
In contrast, the i.i.d.\ sampler requires a significantly longer time of 36.28 seconds.
Code for reproducing all the results can be accessed at the following link: \href{https://github.com/augustofasano/Dynamic-Probit-EP}{https://github.com/augustofasano/Dynamic-Probit-EP}.

\vspace{-5pt}

\section{Discussion}
In this contribution, we have shown how \EP\ can be effectively used to perform approximate inference for the smoothing distribution in dynamic probit models.
The \EP\ Gaussian approximation of the smoothing distribution  allows estimating functionals of interest with computational times that are orders of magnitude smaller than the ones of exact sampling methods.
We have shown in a financial application that the \EP\ approximate moments come with great accuracy, 
despite being endowed with fewer theoretical guarantees than alternative approximation schemes \cite{fasano2022scalable}.
This is in line with recent literature about models for partially-observed Gaussian variables \cite{anceschi2023bayesian,chopin2017leave,fasano2023efficient}, where \EP\ was empirically shown to lead to accurate  approximations of posterior quantities of interest.
Such results constitute empirical guarantees about the use of \EP\ as an approximate method for Bayesian inference and could motivate future research on further theoretical guarantees.

\FloatBarrier 
%
%

\end{document}